\documentclass[10pt,journal,compsoc]{IEEEtran}

\usepackage[noadjust]{cite}
\usepackage[hyphens]{url}
\usepackage{graphicx} 
\usepackage{multirow} 
\usepackage[none]{hyphenat}
\usepackage{subcaption}

\ifCLASSINFOpdf
\else
\fi

\begin{document}

\title{An Experimental Evaluation of TEE technology Evolution: Benchmarking Transparent Approaches based on SGX, SEV, and TDX}

%\title{Benchmarking the TEE technology Evolution: Transparent Approaches based on SGX, SEV, and TDX}

\author{Luigi~Coppolino,
        Salvatore~D'Antonio,
        Davide~Iasio,
        Giovanni~Mazzeo,
        and~Luigi~Romano
        
\thanks{L. Coppolino, S. D'Antonio, G. Mazzeo, and L. Romano are with University of Naples 'Parthenope'. \\email: {first}.{last}@uniparthenope.it}% <-this % stops a space
\thanks{D. Iasio is with Trust Up srl. \\email: davide.iasio@trustup.it}% <-this % stops a space
\thanks{Manuscript received January 16, 2024}}

\markboth{Paper under review at IEEE Transactions on Dependable and Secure Computing}%
{Shell \MakeLowercase{\textit{et al.}}: Bare Demo of IEEEtran.cls for IEEE Journals}

\IEEEtitleabstractindextext{%
\begin{abstract}

Protection of data-in-use is a key priority, for which Trusted Execution Environment (TEE) technology has unarguably emerged as a -- possibly the most -- promising solution. Multiple server-side TEE offerings have been released over the years, exhibiting substantial differences with respect to several aspects. The first comer was Intel SGX, which featured \textit{Process-based TEE} protection, an efficient yet difficult to use approach. Some  SGX limitations were (partially) overcome by runtimes, notably: \emph{Gramine}, \emph{Scone}, and \emph{Occlum}. A major paradigm shift was later brought by AMD SEV, with \textit{VM-based TEE} protection, which enabled "lift-and-shift" deployment of legacy applications. This new paradigm has been implemented by Intel only recently, in TDX.
While the threat model of the aforementioned TEE solutions has been widely discussed, a thorough performance comparison is still lacking in the literature. This paper provides a comparative evaluation of \textit{TDX}, \textit{SEV}, \textit{Gramine-SGX}, and \textit{Occlum-SGX}. We study computational overhead and resource usage, under different operational scenarios and using a diverse suite of legacy applications. By doing so, we provide a reliable performance assessment under realistic conditions. We explicitly emphasize that -- at the time of writing -- TDX was not yet available to the public. Thus, the evaluation of TDX is a unique feature of this study.

\end{abstract}

}

\maketitle

% \fi
\IEEEpeerreviewmaketitle

\section{Introduction}
\begin{figure}
    \centering
    \includegraphics[scale=0.45]{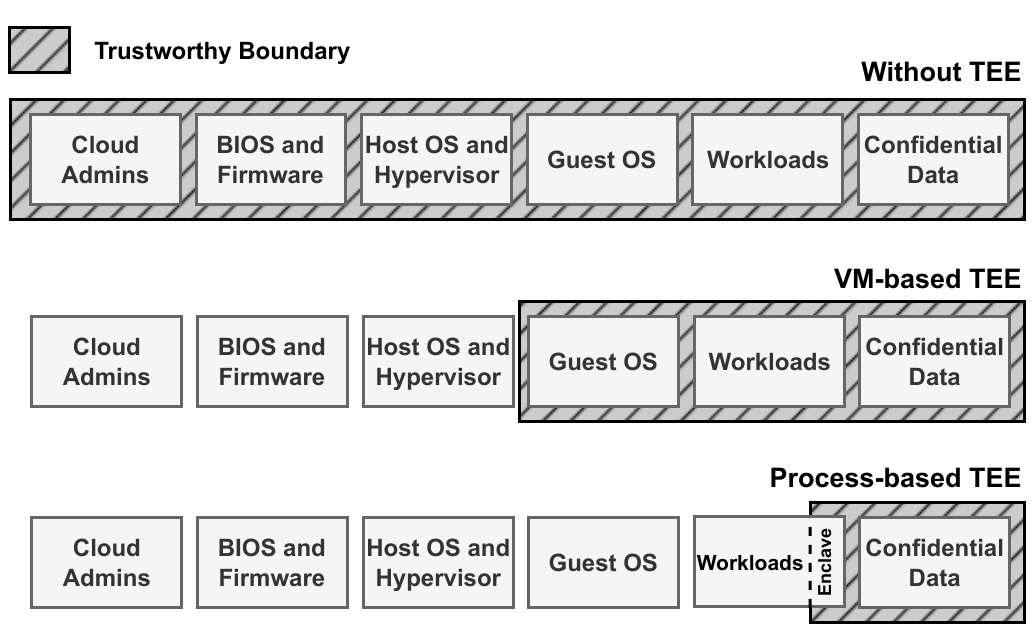}
    \caption{Trust boundaries of current TEE offerings}
    \label{fig:trustboundaries}
\end{figure}

\IEEEPARstart{T}rusted Execution Environments (TEEs) have attracted increasing attention in the quest for secure computing, largely because this technology has much better performance than alternative solutions, such as Homomorphic Encryption or Secure Multi-Party Computation \cite{survey}. Protection of data-in-use in untrusted cloud computing platforms was initially enabled by \textit{Process-based TEE} solutions, which relied on Intel \textit{Software Guard eXtensions} (SGX) \cite{sgx}. Working with SGX, researchers and practitioners from the academia and the industry identified drawbacks which limited the applicability of this technology. Major concerns were related to memory constraints and programming restrictions, which made the adaption of legacy software to SGX not only challenging but also prone to errors. The enrichment of the Intel SGX technology with a runtime layer --- e.g., \textit{Gramine} \cite{gramine} (formerly known as \emph{Graphene}), \textit{Occlum} \cite{occlum}, or \emph{Scone} \cite{scone} --- helped to mitigate porting issues but at the cost of a larger Trusted Computing Base (TCB).
AMD with the \emph{Secure Encrypted Virtualization} (SEV) \cite{sev} technology introduced the concept of a \textit{VM-based TEE} (also known as \textit{Confidential VMs}), which is significantly more user-friendly, since it allows existing applications to run in the secure environment without any modification. The downside, as compared to Process-based TEE, is a weaker threat model. A VM-based TEE has been recently presented by Intel, too. It is called \textit{Trust Domain eXtension} (TDX) \cite{tdx}, and builds on lessons learned from SGX. 
\\While the threat models of these technologies (see Figure \ref{fig:trustboundaries}) are well known, and detailed analyses of the tradeoffs of alternative solutions have been made \cite{sahita}\cite{felicitas}\cite{fei}, the scientific/technical literature provides limited coverage of performance evaluation of TEE offerings, since the currently available comparison of existing TEE approaches for transparent --- or quasi-transparent --- protection of data-in-use from a quantitative point of view is largely incomplete. This undermines the possibility for security engineers/researchers to take informed decisions about the specific TEE solution to use, based on individual application requirements. There are some previous research works featuring  comparative analyses of TEE solutions \cite{gottel-sevsgx}\cite{akram-sevsgx}\cite{mofrad-sevsgx}, which mainly focused on setting side by side SGX and SEV. In just one case, \textit{Gramine-SGX} was also included in the evaluation. No experimental evaluation exists to date on TDX, since this technology has only recently been released and --- at the time of this writing --- there are still no commercial servers available on the market equipped with TDX technology (and additionally the Linux kernel still lacks stable TDX support). 

In this work, we delve into a comprehensive comparative analysis of a wide spectrum of solutions for transparent TEE support, ranging from earlier proposals (namely: \textit{Gramine-SGX} and \textit{Occlum-SGX}) to the most recent one (namely: \textit{TDX}). Notably, we are the first ones to provide an experimental evaluation of TDX (as already mentioned, TDX is not publicly available yet, but we were granted complimentary access to a research instance of an Intel TDX powered machine, which gave us the possibility of running our experiments). We investigate the performance tradeoffs of alternative TEE solutions, with respect to the deployment of legacy applications. Importantly, the study meticulously evaluates performance metrics -- such as computational overhead and CPU utilization -- which are crucial in understanding the practical implications of deploying applications on TEE solutions in real-world scenarios (including the costs of the cloud platform setup). We selected a diverse set of legacy applications, which  collectively make for a substantial benchmark suite, representing typical use cases. By doing so, we are able to provide a realistic and comprehensive assessment of each TEE approach. More precisely, the experimental activity focuses on workloads with different resource usage profiles: \emph{i}) CPU-intensive - \emph{TensorFlow} and \emph{Pytorch}; \emph{ii}) Memory-intensive - \emph{Redis} and \emph{Hashicorp Vault}; \emph{iii}) I/O-intensive - \emph{NGINX} and \emph{NodeJS}. By thoroughly evaluating the complexity of integration issues and the performance trade-offs of alternative TEE solutions -- covering both process-based and VM-based proposals -- our study provides practitioners with a compass for navigating in the  challenging space of effectively using these technologies for security improvement of legacy software. 
% Our findings aim to guide stakeholders in navigating the complexities of TEE integration.
\\The experimental campaign produced the following outcomes: 
\begin{itemize}
\item VM-based TEEs are faster (i.e. they have smaller execution times),  particularly when handling memory- and I/O- intensive workloads, as compared to Process-based TEEs. They are also characterized by a lower overhead in terms of resource usage. 
\item Although less performing than VM-based TEEs, the overhead of Process-based TEEs is lower in the case of CPU-intensive workloads (as opposed to memory- and I/O- intensive ones). Since the trust model of Process-based TEEs is stronger, Process-based TEE solutions can thus be the right choice for these workloads, because in many setups they represent a good compromise between performance penalty and security improvement.  
\item In the domain of VM-based TEEs, TDX outperforms SEV in terms of efficiency. We explicitly note that the performance gap between the two security solutions is much larger than performance gap between the respective CPUs.
\item In the domain of process-based TEEs, \textit{Gramine-SGX} consistently outshines \textit{Occlum-SGX}, not only in performance but also in terms of resource consumption.
\end{itemize}
The remainder of this work is organized as follows. Section \ref{background} gives a background on the TEE solutions and defines the motivation behind our paper. Section \ref{relwork} presents previous works focused on the evaluation of the TEE approaches that we cover in this paper. Section \ref{methodology} describes the evaluation methodology and defines the setup used to conduct the experiments. Section \ref{analysis} discusses the outcomes of the experimental campaign. Section \ref{servicecost} reports an analysis of the impact of TEE solutions on Cloud service costs. Finally, Section \ref{conclusion} provides a summary of and comments the main findings.

\section{Background and Motivation}
\label{background}
\begin{figure}
    \centering
    \includegraphics[scale=0.55]{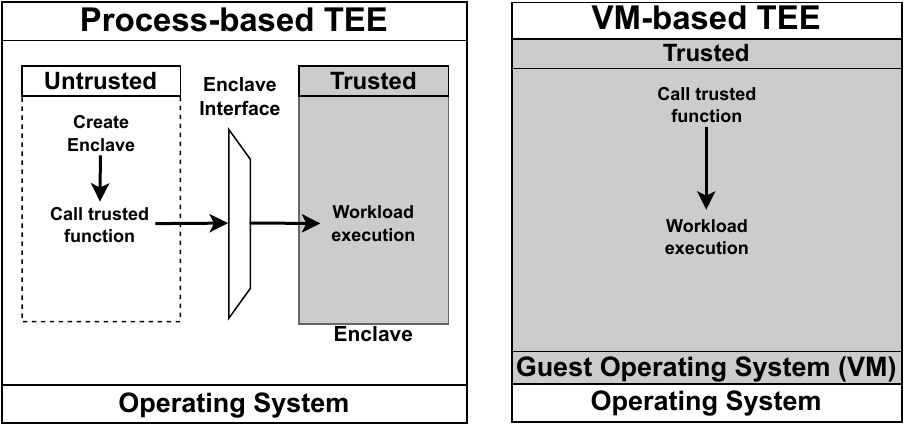}
    \caption{Process-based vs VM-based TEE}
    \label{fig:tee}
\end{figure}

In the domain of confidential computing, there is a clear-cut division between Process- and VM- based Trusted Execution Environment (TEE) (Figure \ref{fig:tee}), whose common goal is ensuring security of data-in-use. In this section, we overview the technologies that are under the magnifying glass of this experimental work. Moreover, we present key aspects that motivate our paper. 

\subsection{Process-based TEE}
In Process-based TEEs, a process is split into two parts: one considered secure (trusted) and the other considered not secure (untrusted). The secure part is located in encrypted memory, managing sensitive computations, while the non-secure part communicates with the operating system and moves I/O from the encrypted memory to other parts of the system. Data transfer into and out of this secure memory zone is tightly regulated, with stringent controls over the data size and type that is allowed to cross the boundaries. Ideally, data transferred to or from the encrypted memory should be encrypted during transit and only decrypted within the TEE, ensuring that it is only accessible to the software operating within the TEE. The widely adopted Process-based TEE for server-side security is Intel Software Guard eXtension (SGX) \cite{sgx}. 
\\SGX enables the creation of \textit{secure enclaves} within the processor, isolating sensitive code and data from the rest of the system thanks to the extension of Intel’s Instruction Set Architecture (ISA) by 18 new instructions. Sensitive code and data are stored in the \textit{Enclave Page Cache} (EPC), a specific 128 MiB / 256 MiB (for SGX v1 or v2) section of memory set aside during the system startup for storing the code and data of enclaves. Any attempt to access an enclave's page outside of the EPC results in a page fault. The SGX driver collaborates with the CPU to determine which pages to remove from the cache. The memory encryption engine (MEE) ensures that communication between the CPU and system memory remains secure, and it is also responsible for preventing tampering and providing replay protection. An enclave can only run in user mode (\emph{ring3}) since the host OS is considered untrusted. This means that \emph{system calls} cannot be invoked from inside the TEE. The only way to execute them is through well-defined interface calls outside the enclave. An important SGX feature for verifying the integrity and security of an enclave is the attestation mechanism. SGX supports a local attestation, used for communication between enclaves on the same platform, and a remote attestation used for demonstrating trustworthiness to external entities. 

Developing applications for SGX can be challenging because the application must be refactored into trusted (within the enclave) and untrusted components. This porting requires careful design to ensure security and can be a complex and time-consuming process.
This is where SGX runtimes came to the rescue. They act as intermediate layers that abstract away the complexity of SGX and allow developers to run applications within SGX enclaves with much fewer changes.
By using these runtime environments, developers can more easily take advantage of the security benefits of SGX, allowing sensitive or critical applications to be deployed in potentially untrusted environments.
\\The common approach adopted by these runtimes to enable quasi-transparent porting is to execute system calls directly inside the enclave via a \textit{Library OS (LibOS)}, i.e., a new paradigm trend where kernel functions are available to user space (\textit{ring3}) programs in a form of a library.
There are several runtime environments for Intel SGX (e.g. \textit{SCONE, Gramine, Occlum, SGX-LKL}).
In this work, we focus on \textit{Gramine} and \textit{Occlum} because they stand out among the open-source solutions as the widely adopted ones \cite{ccconsortium}. 

\subsubsection{Gramine-SGX}
Gramine \cite{gramine} is a runtime coming with a lightweight library OS, which facilitates the use of dynamically loaded libraries and runtime linking. It stands out as one of the runtimes that fully accommodate fork/clone/execv system calls, which are essential for multi-process abstraction, thereby supporting a wide spectrum of applications. A distinctive attribute of Gramine \cite{gramine} is its ability to secure dynamic loading, allowing users to incorporate any library into an enclave while ensuring the integrity of the libraries. It enables the safe execution of any binaries, such as those using \textit{glibc} with dynamically linked libraries within the enclave. To do this, a user of Gramine must create a \textit{manifest} detailing all the trusted libraries and data files employed within an enclave and then sign this manifest to safeguard its integrity before executing the chosen binary on SGX. Gramine provides a basic set of system calls in its capacity as a LibOS, which can be processed rapidly due to the low interaction with the host OS. Alternatively, system calls that are not supported by the library OS are meticulously handed over to the host OS. This handover necessitates exits from and re-entries into the enclave, leading to substantial performance costs. Moreover, because the host OS is considered untrustworthy in the SGX security framework, Gramine also verifies the host OS's responses. Hence, system calls that are passed to the host OS incur greater costs compared to those that the library OS can emulate.

\subsubsection{Occlum-SGX}
Occlum \cite{occlum}, while sharing similarities with Gramine as a runtime environment, owes its spread and adoption to its strong community support. Being an open-source project, Occlum benefits from contributions from a range of developers, which helps in its development and maintenance. This community involvement can be important for ensuring the tool stays updated and relevant.
This system introduces Software Fault Isolation-Isolated Processes (SIPs) within a LibOS in an enclave's single address space.
Software Fault Isolation (SFI) is a technique for sandboxing untrusted modules in different domains.
The novel aspect of this proposal is the \textit{Memory Protection Extensions-based (MPX), Multi-Domain SFI (MMDSFI)}, which supports an unlimited number of domains without restrictions on their addresses and sizes.
This allows for enhanced intra-enclave isolation, including isolation between processes and between a process and the LibOS. To ensure the security and compliance of these isolation mechanisms, an independent binary verifier called the \textit{Occlum verifier} is introduced. This verifier statically checks ELF binaries to ensure they adhere to MMDSFI's security policies.

\subsection{VM-based TEE}
VM-based TEEs foresee that an entire VM memory is encrypted using keys sealed in the hardware, which prevent interference by a malicious VMM. Current technologies such as Intel TDX and AMD SEV provide dedicated ephemeral encryption keys for each VM, thus also protecting the VMs from each other.

\subsubsection{AMD SEV}
AMD SEV (Secure Encrypted Virtualization) \cite{sev} is a security feature in AMD EPYC processors, which utilizes AMD Secure Memory Encryption (SME) and AMD Virtualization (AMD-V) for cryptographically separating VMs from the hypervisor. Each VM gets a distinct, temporary AES key for encrypting memory during operation. The AES mechanism in the processor's memory controller handles encryption and decryption of data to and from the main memory. These keys for each VM are overseen by the AMD Platform Security Processor (PSP). A specific bit (C-bit) in physical addresses is used to encrypt memory pages. SEV also offers remote attestation, enabling VM owners to check the integrity of VMs and the SEV platforms. The PSP creates an attestation report, signed by an AMD-certified key, which VM owners can authenticate along with platform and guest measurements. AMD has introduced three versions of SEV: the first only secures VM memory confidentiality; the second, SEV-ES (Encrypted State), additionally safeguards CPU register states during transitions with the hypervisor; the third, SEV-SNP (Secure Nested Paging), further protects against memory attacks like corruption, replaying, and remapping.

\subsubsection{Intel TDX}
Intel Trust Domain Extensions (TDX) \cite{tdx}, as part of the 4th Generation Intel Xeon Scalable Processor, is built using a combination of Intel Virtual Machine
Extensions (VMX) ISA extensions, multi-key, total memory-encryption (MKTME) technology, and a CPU-attested, software module. In addition to these technologies, TDX leverages the Intel SGX for what concerns the attestation of Trusted Domains (TDs). 
Intel TDX enhances the security of TDs by offering protection against certain types of attacks that involve physical access to platform memory. This includes protection against offline attacks, like DRAM analysis, which encompasses cold-boot attacks, as well as active attacks on DRAM interfaces. These active attacks might involve intercepting, altering, moving, splicing, or creating aliases for memory contents. However, Intel TDX does not provide a defense against the replay of memory content through physical attacks. Confidentiality and Integrity of Memory and CPU state is achieved by excluding elements such as firmware, software, devices, and cloud platform operators from the trusted computing base (TCB), giving workloads more secure access to CPU instructions and other technologies. This capability is independent of the cloud infrastructure used to deploy the workload.
Remote attestation is another feature provided by TDX, enabling the validation of a workload's environment and the security integrity of the TCB.

\subsection{Motivation}
\label{sec:motivation}
As the commercial offering of TEE has expanded over the years, it has become challenging for security engineers and decision-makers to select the right solution matching their requirements. Understanding the performance implications of TEEs is essential. This necessity stems from the need to balance security features with system efficiency, ensuring that the implementation of TEEs does not hinder system performance. Performance metrics such as computational overhead, resource utilization, and impact on response times are critical in determining the viability and appropriateness of TEEs in various operational contexts. Comprehensive knowledge of these aspects enables informed decisions about deploying, configuring, and optimizing TEEs, thus ensuring robust security without compromising on performance. Furthermore, the higher resource usage given by TEEs also entail higher expenses for cloud deployments. Overall, the decision is taken by considering the following questions:
\begin{itemize}
    \item \emph{How does it impact the performance of the application?}
    \item \emph{How does it impact the isolation of sensitive data?}
    \item \emph{How does it impact infrastructural costs?}
    \item \emph{How does it impact the personnel costs?}
    \item \emph{How does it impact the migration effort?}
    \item \emph{How does it impact the availability of the application?}
    \item \emph{How does it impact the customer security perception?}
\end{itemize}
Given the diversity of solutions, we believe it is important to provide an insight into the performance of the most popular approaches. Each approach has unique characteristics in terms of design, operation, and performance implications. A comparative analysis is essential to understand the trade-offs and benefits of each method.
Different applications may have varying requirements. For instance, a blockchain application might prioritize integrity and isolation, a fin-tech application might focus on performance, and a critical infrastructure application might wonder about reliability. A comparative research can help stakeholders select the most appropriate TEE approach for their specific use case. 
\\The balance between security and usability is an age-old challenge. Transparent security aims to minimize user friction while maximizing protection. Our research work can contribute to designing TEEs that better align with user expectations and application requirements.
Last but not least, with regulations like GDPR and CCPA imposing stringent data protection requirements, understanding the performance implications of TEEs can assist organizations in making informed decisions that comply with legal standards.

\section{Related Work}
\label{relwork}
In this section, we report previous research works that experimentally evaluated TEE technologies using different categories of workloads. 
\\Akram et al. \cite{akram-sevsgx} analyzed the overhead and memory layout issues of Intel SGX and AMD SEV. They chose conventional scientific computing workloads in conjunction with advanced applications that meet the criteria of the High-Performance Computing (HPC) application space. Their assessment included workloads traditionally utilized to benchmark HPC frameworks, particularly the NAS Parallel Benchmark (NPB) suite. This suite, comprising different kernels and pseudo-applications, has been a long-standing tool for examining HPC frameworks. In addition to traditional scientific computing, they also put attention on machine learning, graph analytics, and emerging scientific computing
workloads. For all the evaluations, they conducted tests without hyperthreading by restricting the number of threads to the number of cores on each platform. Regarding SGX, programs were compiled statically and connected against a modified standard C library in SCONE. With SEV, instead, they utilized AMD-provided scripts to set up the SEV-enabled host machine and the guest virtual machine managed by QEMU.

Gottel et al. \cite{gottel-sevsgx} provided useful insights into the energy, latency, throughput, and processing time of AMD SEV and Intel SGX. In their study, authors analyzed these two technologies within the context of large-scale distributed systems operating on sensitive data within public cloud infrastructures. The porting of the workload in SGX was realized using \textit{Gramine-SGX}. 
The authors explained the differences and similarities, and threat models, of the SGX and SEV hardware
architectures. They discussed also the engineering efforts in adopting both Intel and AMD hardware solutions (individually). The performance evaluation was conducted on SGX and SEV using memory-intensive micro-benchmarks. Specifically, they executed an evaluation study leveraging a complete prototype of an event-based publish/subscribe system. Finally, they deployed a realistic scenario and workloads over a publish/subscribe implementation to gather experimental data in real-world settings. Moreover, the authors recorded the power consumption of the publish/subscribe system to identify how it varies based on the adopted technology. 

Mofrad et al. \cite{mofrad-sevsgx} also compared Intel SGX and AMD SEV, emphasizing their functionality, use cases, security attributes, and performance consequences. The authors provided information about the characteristics and application scenarios of these technologies.
They investigated the design architecture and attack surface of Intel SGX and AMD Memory Encryption technologies. To accomplish their benchmarks, they crafted various applications compatible with both SGX and AMD benchmarks, employing standard C/C++ library functions for a uniform code base and an equitable benchmarking environment. Their focus lies in assessing the performance of the Intel and AMD Memory Encryption Engine and other architectural components when operated under similar code base conditions.
Their benchmarks are segmented into three distinct categories: the first evaluates the TEE's capacity for executing intensive floating-point operations without data wrangling; the second assesses the Memory Encryption Engine's performance of both TEEs through data sorting; and the third inspects the overall performance of TEEs within a security protocol in a complex application used in public cloud environments.

Our paper stands out from previous works especially for the evaluation of the new Intel TDX technology. It introduces a novel perspective in the context of TEE research, evaluating the entire spectrum of current approaches for near-transparent porting of applications into TEEs. We conduct a comprehensive comparison of the performance across Intel TDX, AMD SEV, Gramine, and Occlum. Moreover, our work uses a wide set of workloads characterized by completely different resource usage profiles. 

\section{Methodology}
\label{methodology}
In the following, we outline the hardware and software settings utilized for the experiments, which are analyzed in the rest of the paper.

\subsection{Environment}
\begin{figure}
    \centering
    \includegraphics[scale=0.45]{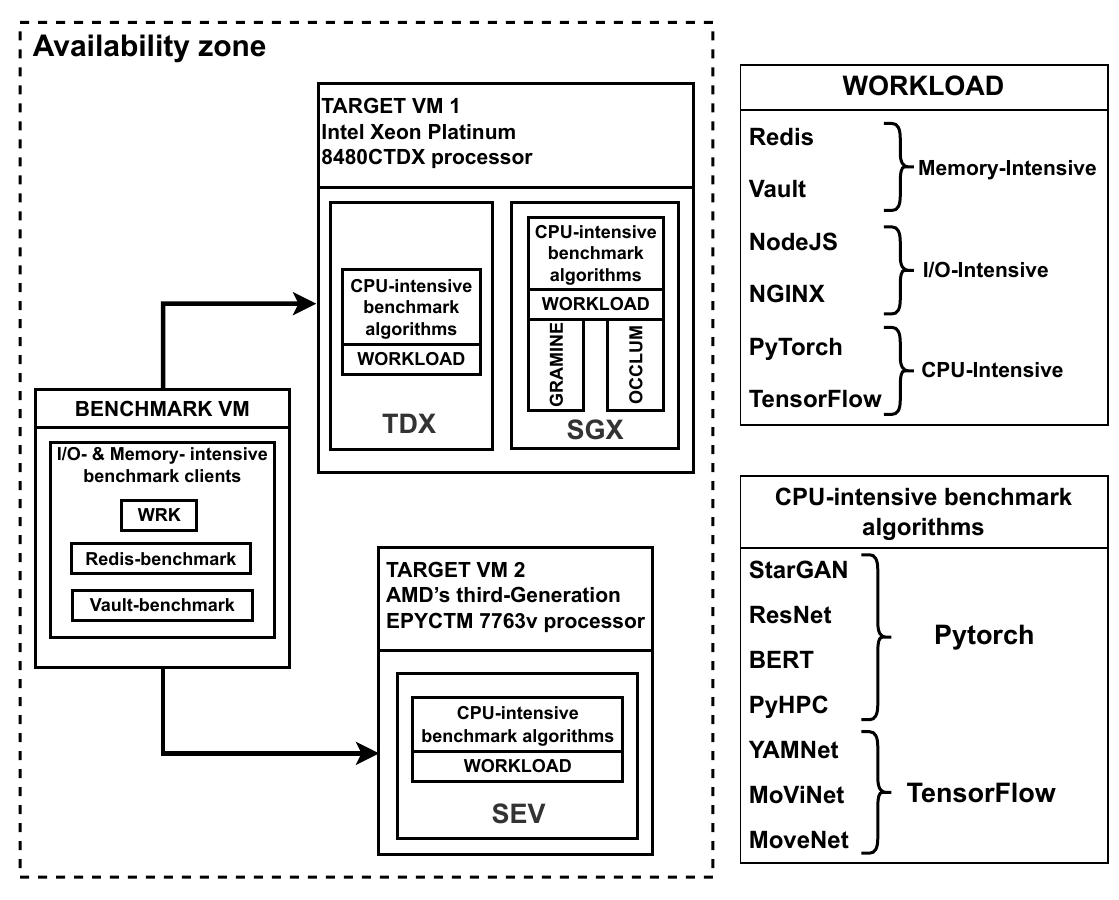}
    \caption{Experimental Setup}
    \label{fig:expsetup}
\end{figure}

Figure \ref{fig:expsetup} shows our experimental setup. In order to get results as comparable as possible, we configured our machines hosting the workloads with the same amount of virtual CPUs (vCPUs) cores (4 cores), the same amount of virtual RAM (vRAM) (16GB), and similar NIC characteristics. 
\\If the workload requires to be stimulated in a client/server topology, there is the need for an external benchmark client to do requests to the server. In this case, in order to prevent interference, a separate VM was deployed. Low-latency channels between the client and server are important to get fair results. Hence, the benchmark VM was deployed in the same \emph{Availability Zone} as the workload VM.
\\At the time of writing this paper, there is no public availability of an Intel TDX machine. Experiments on this technology were conducted using a server offered by Intel for research purposes, which mounts an Intel Xeon Platinum 8480CTDX ($2.0GHz$, Turbo at $3.8GHz$). Well-known virtualization tools like QEMU and libvirt are needed for Intel TDX to enhance the confidentiality of active workloads. For the effective operation of a confidential VM, various elements within the virtualization stack must be compatible with TDX hardware. Intel is actively engaged in integrating comprehensive TDX software support into the upstream versions of the Linux kernel, QEMU, and libvirt. We leveraged the patched versions of Linux kernel, QEMU, and libvirt to deploy a guest TDX VM. Even in this case, the same amount of cores and memory are configured on the mounted QEMU Confidential VM. 
\\The TDX machine embeds the SGX extension as well. So, we decided to run SGX (i.e., Gramine and Occlum) and \emph{Native} runs  -- useful to set the baseline for the comparison -- in the TDX server. In this way, we can provide a fair comparison of Native, TDX, and SGX.  
At the same time, we also want to provide information on how SGX performance compares between the old and new generations of CPUs. Hence, we instantiated a Standard DCsv3-series VM, which uses the 3rd Generation Intel Xeon Scalable (Icelake) 8370C ($2.9GHz$, Turbo at $3.5GHz$) with the SGXv2 capabilities (i.e., larger EPC and support for dynamic memory allocation (EDMM)). 
\\Experiments on AMD SEV leveraged a Standard DCasv5-series VM, which uses AMD's third-Generation EPYCTM 7763v processor ($2.5GHz$, Turbo at $3.5GHz$). We selected this server because it is one of the best-performing SEV-enabled AMD machine, which is comparable with the Intel machine. 
\\In terms of software, all VMs were configured with Ubuntu 22.04.3 LTS. We use the latest versions of \textit{Gramine-SGX} (v1.6), and Occlum (v0.30.0). The SGX SDK software stack relies on version 2.22. 

\subsection{Managing CPUs with different Clock Frequencies}
In our experimental setup, we make an identical configuration of vCPUs and vRAM across the different nodes. However, there are still some factors that could potentially skew the fairness of our results such as the range of CPU clock frequencies. The clock frequency varies slightly among the nodes we selected for our experiments. This variation is significant enough that it must be considered when analyzing our experimental results to ensure accuracy and fairness.
\\To address this potential discrepancy, we adopt a method of normalization for our calculated overheads. This normalization process involves using microbenchmark results that are publicly available for the specific CPUs used in our servers \cite{8375bench}\cite{8480bench}\cite{amdbench}. These benchmarks provide a detailed analysis of the performance capabilities of these CPUs under various conditions. By incorporating these benchmark results into our analysis, we can adjust our data to account for the differences in clock frequency. This normalization allows us to compare performance metrics more accurately across different hardware setups. It ensures that any observed differences in the experimental results are due to the factors we are testing, rather than inherent differences in the hardware's basic processing speed. In essence, this approach helps us isolate the variables we are interested in studying, by mitigating the impact of an external variable – the clock frequency range – that could otherwise introduce an element of unfairness into our results. This careful consideration and adjustment for hardware differences underscore the rigor and precision we are applying in our experimental analysis.

\subsection{Workloads}
To ensure that our evaluation of Trusted Execution Environment (TEE) approaches is thorough and reflective of real-world scenarios, it is crucial to test them across a diverse range of workloads. These workloads should mirror the variety of applications commonly used in practice, each with its unique resource utilization characteristics. We have carefully selected several prevalent workloads, categorized based on their primary resource demands: CPU-intensive, memory-intensive, and I/O-intensive.

\par{\textit{CPU-Intensive Workloads.}}
For tasks that demand substantial CPU resources, particularly for computation-intensive processes, we selected the following workloads:

\begin{itemize}
\item \textit{TensorFlow}\footnote{\url{https://www.tensorflow.org}}: This is a widely used framework in the field of deep learning. We specifically focused on running machine learning inference algorithms, which are known for their high CPU usage due to complex calculations.
\item \textit{PyTorch}\footnote{\url{https://pytorch.org}}: Another deep learning framework, PyTorch is renowned for its efficiency in performing computations that require significant CPU power.
\end{itemize}

\par{\textit{Memory-intensive Workloads.}} For workloads that predominantly consume memory resources, we included:

\begin{itemize}
\item \textit{Redis}\footnote{\url{https://redis.io/}}: Known for its efficiency as an in-memory key-value store, we utilized Redis for operations like deep in-memory scanning, and typical SET and GET commands, which are memory-intensive.
\item \textit{Hashicorp Vault}\footnote{\url{https://www.hashicorp.com}}: This tool is used for key and secrets management and is known to be memory-intensive, making it a suitable test for our memory workload category.
\end{itemize}

\par{\textit{I/O-intensive Workloads.}}
To evaluate the performance under I/O stress, we selected:

\begin{itemize}
\item \textit{NGINX}\footnote{\url{https://nginx.org}}: A high-performance web server, NGINX is used for serving web content, which is typically I/O-intensive due to the nature of web traffic and data transfer.

\item \textit{NodeJS}\footnote{\url{https://nodejs.org}}: Known for server-side scripting, NodeJS applications often involve significant I/O operations, especially when handling multiple concurrent requests.
\end{itemize}

For each of these workloads, we conducted multiple runs to ensure reliability and accuracy in our results. Specifically, we set a confidence interval of $95\%$ and empirically verified that 10 repetitions of our experiments were enough to achieve the aforementioned target. The outcomes were averaged to account for any anomalies and to provide a more accurate representation of the performance. This rigorous testing methodology allows us to comprehensively assess the effectiveness of TEE approaches across a spectrum of real-world applications, ensuring that our findings are both valuable and applicable to a wide range of scenarios.

\subsection{Benchmarks}
We employed a variety of benchmarking tools to stimulate the different workloads. It is important to notice that their final configuration --- e.g., the number of connections, and the ranges of parallel clients --- was obtained empirically after several preliminary experiments aimed at reaching the workload saturation point.

\par{\emph{Redis}}. We used \emph{redis-benchmark}, a tool specifically designed for the REDIS key-value store. It is used to measure the performance of a Redis server by running a series of predefined tests. In our study, we used redis-benchmark to make typical operations of SET and GET, thus evaluating the throughput and latency of the \textit{Redis} server during writing and reading from memory. In terms of configuration, we kept the total number of connections to a fixed value of $100k$ and varied the number of parallel clients from $10$ to $1000$. 
\par{\emph{NGINX \& NodeJS}}. The \emph{wrk2} benchmark was adopted to generate a significant load against NGINX and NodeJS. It provides a flexible scripting interface that allows us to simulate different types of HTTP requests, which is crucial for testing the performance of NGINX as a web server and NodeJS in server-side scripting scenarios. By adjusting the number of connections, threads, and the duration of the test, we were able to assess how these servers handle high traffic and concurrent connections. The benchmark was configured with a duration of $30s$, a fixed number of connections to $100$, and a varying number of clients ranging from $10$ to $16000$.  
\par{\emph{Vault}}. The \emph{vault-benchmark} tool --- specifically designed for Hashicorp Vault --- helps in evaluating the performance of Vault in various scenarios, including reading and writing secrets, authentication requests, and other secret management operations. We used the vault-benchmark to determine the throughput and response times of Vault during \textit{static\_secret\_writes} operations, which is critical for understanding its scalability and reliability in a production environment. The benchmark was tuned with $numkvs=100$, and a $kvsize$ varying in the range $[10,800]$.
\par{\emph{PyTorch}}. The benchmarking of PyTorch involved the adoption of built-in algorithms, which range from image processing and classification to natural language processing and high-performance computing simulations. Specifically:
\begin{itemize}
    \item \textit{StarGAN} (\textit{pytorch\_stargan-cpu}) - Image-to-image translations.
    \item \textit{ResNet} (\textit{phlippe\_resnet-cpu}) - Image classification.
    \item \textit{BERT} (\textit{BERT\_pytorch-cpu}) - Natural language processing.
    \item \textit{PyHPC} (\textit{pyhpc\_isoneutral\_mixing-cpu}) - Scientific simulations in fluid dynamics and climate modeling.
\end{itemize}

\par{\emph{TensorFlow}}. Even in this case, we used built-in models to evaluate performance:

\begin{itemize}
    \item \textit{YAMNet} - a deep learning model designed for audio event detection and classification. 
    \item \textit{MoViNet} (movinet\_stream) - a family of models optimized for video understanding, particularly for streaming video analysis. 
    \item \textit{MoveNet} - a cutting-edge model for human pose estimation, known for its speed and accuracy. It is designed to be lightweight and efficient, making it suitable for real-time applications.
\end{itemize}

\section{Experimental Results}
\label{analysis}
In this section, we dive into the analysis of results obtained during our experimental campaign. The focus is on interpreting the data collected, evaluating the outcomes against our hypotheses, and understanding the implications of these findings. For CPU-Intensive workloads, we compare the execution time among the different technologies. For Memory and I/O-intensive workloads, instead, we analyze the throughput and latency. In all cases, we also report details of an analysis of the average CPU utilization. 
\\All graphs are normalized as follows:

\[x_{norm}=(x-x_{min})/(x_{max}-x_{min})\]

\begin{figure}
    \centering
    \includegraphics[scale=0.35]{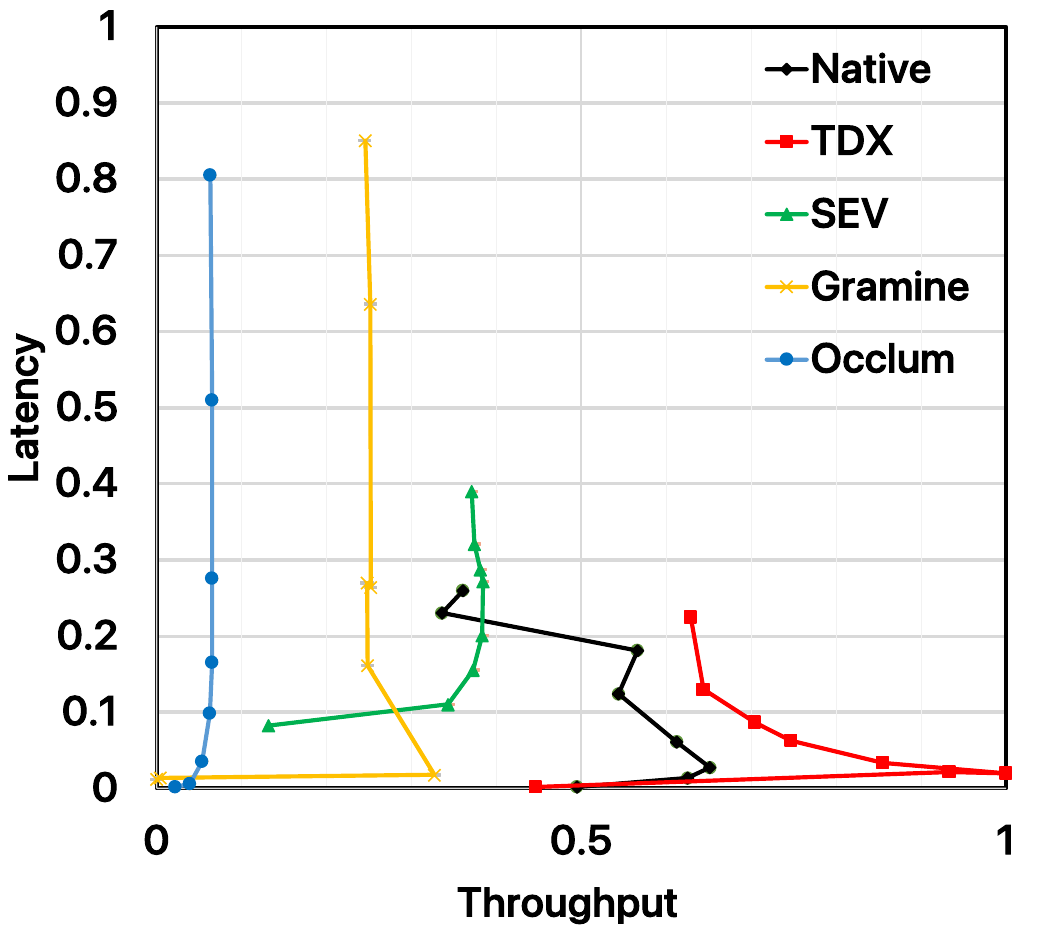}
    \caption{Redis Performance -- Throughput vs Latency}
    \label{fig:redis-perf}
\end{figure}

\begin{figure}
    \centering
    \includegraphics[scale=0.35]{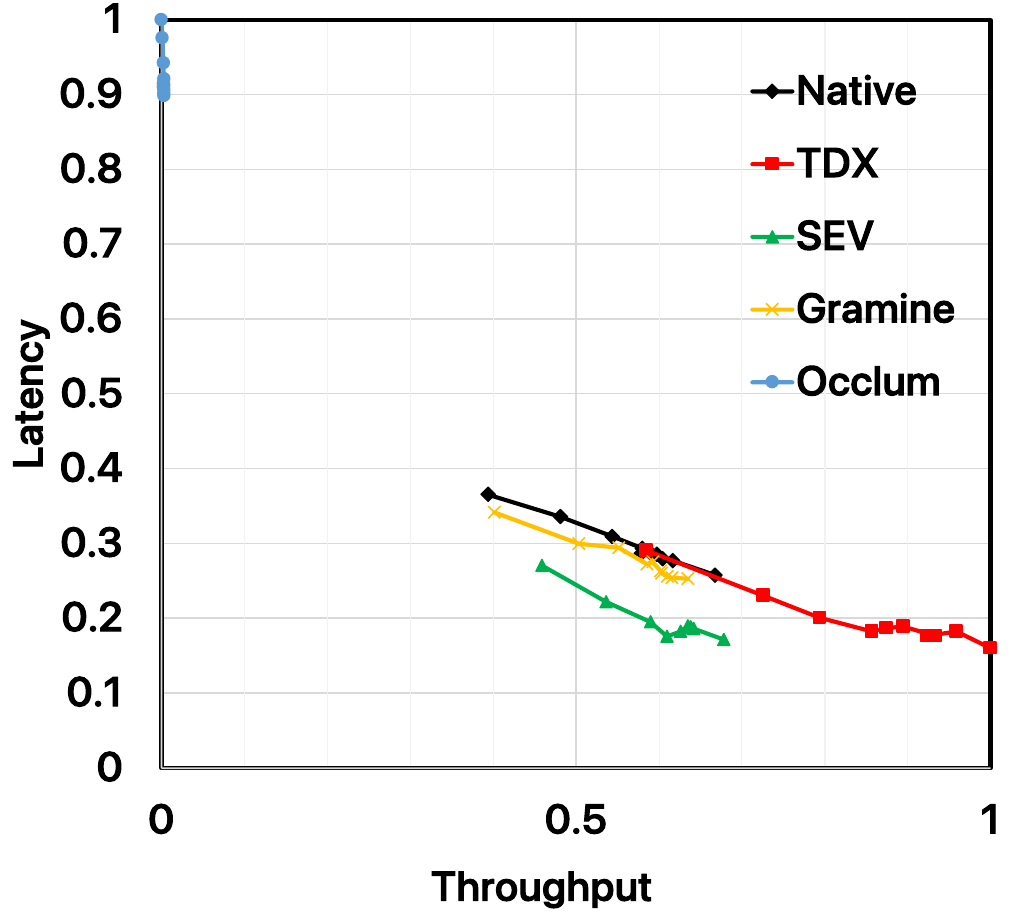}
    \caption{Vault Performance -- Throughput vs Latency}
    \label{fig:vault-perf}
\end{figure}

\subsection{Memory-Intensive Workloads}
Figure \ref{fig:redis-perf} shows the graph on \textit{Redis} performance. The $x$-axis refers to the throughput and the $y$-axis to the latency. As expected, \textit{Gramine} (yellow line with 'X' marker) and \textit{Occlum} (blue line with star marker) have the worst performance. It can be observed they reported the highest latency and low throughput, although \textit{Gramine} performs better than \textit{Occlum}, which experiences a sharp peak, and then a rapid decrease, suggesting inefficiency, especially at lower throughputs. The \textit{TDX} solution (red line with square marker) has surprisingly the highest throughput and a latency that is comparable with the one observed on the \textit{Native} (black line with circle marker). The \emph{Native} solution has a steady increase in latency with high throughput. Furthermore, we can notice that \textit{Native} and \textit{TDX} have a similar trend. 
\textit{SEV} (green line with triangle marker) has higher latency at the beginning, which sharply increases with a small increase in throughput, indicating a potential bottleneck in handling higher loads. However, it is important to notice that the difference in performance between TDX and SEV depends on two factors: the CPU typology and the security technology itself. The impact of the CPU typology can be obtained using publicly available benchmarks \cite{8375bench}\cite{amdbench}, which tell us that the AMD CPU is on average $40.7\%$ slower than the Intel CPU, which hosted the execution of all the other versions of the workloads (Native, TDX, Gramine, Occlum).  If we subtract the $40.7\%$ we can argue that the actual overhead of SEV with respect to TDX can be considered of $\approx22\%$. 
\\In Figure \ref{fig:vault-perf}, we report the \textit{Vault} performance. The rightest points correspond to the lowest \textit{kvsize}. The increase of \textit{kvsize} causes the decrease of the throughput and a rise in the latency. Even in this case, we  observed that TDX provides the highest throughput. It is interesting to notice that SEV has a throughput similar to the \textit{Gramine} and \textit{Native} solutions but at the same time, it has the lowest latency. A different story is Occlum, which experienced very bad performance as can be noticed by the graph highlighting the high latency and the low throughput.

\begin{figure}
    \begin{subfigure}{0.4\textwidth}
        \includegraphics[scale=0.35]{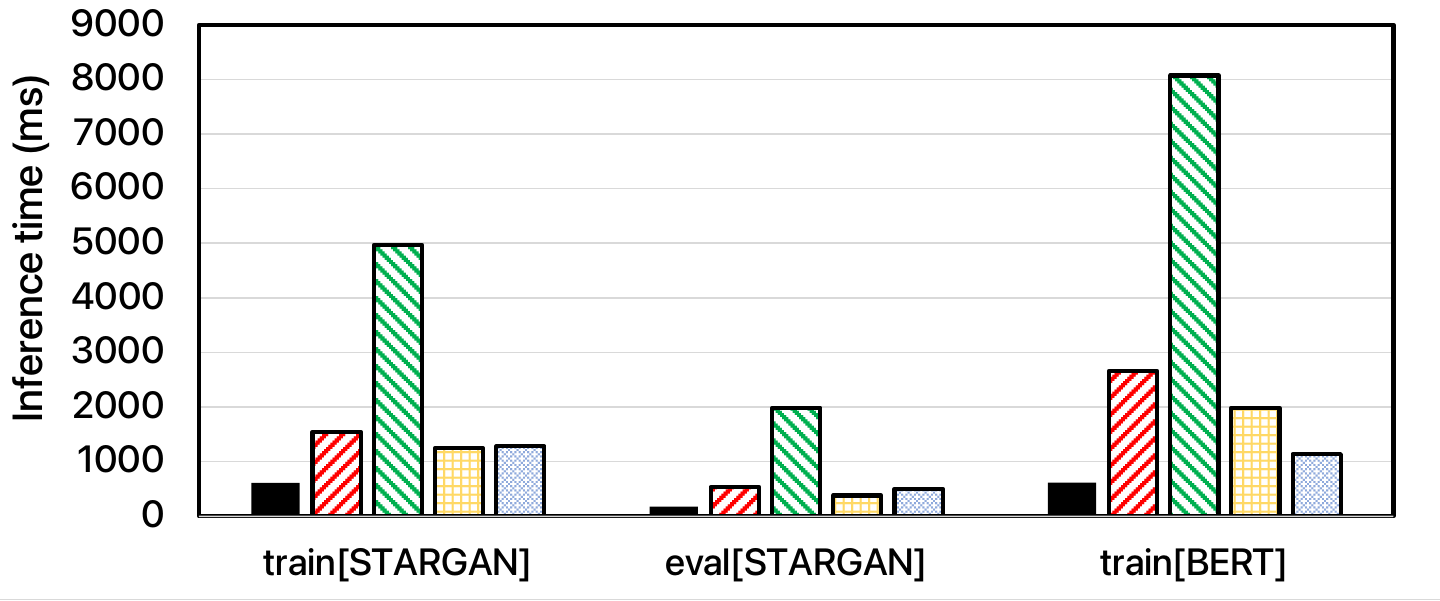}
        \label{fig:first}
    \end{subfigure}
    \hfill
    \begin{subfigure}{0.4\textwidth}
        \includegraphics[scale=0.35]{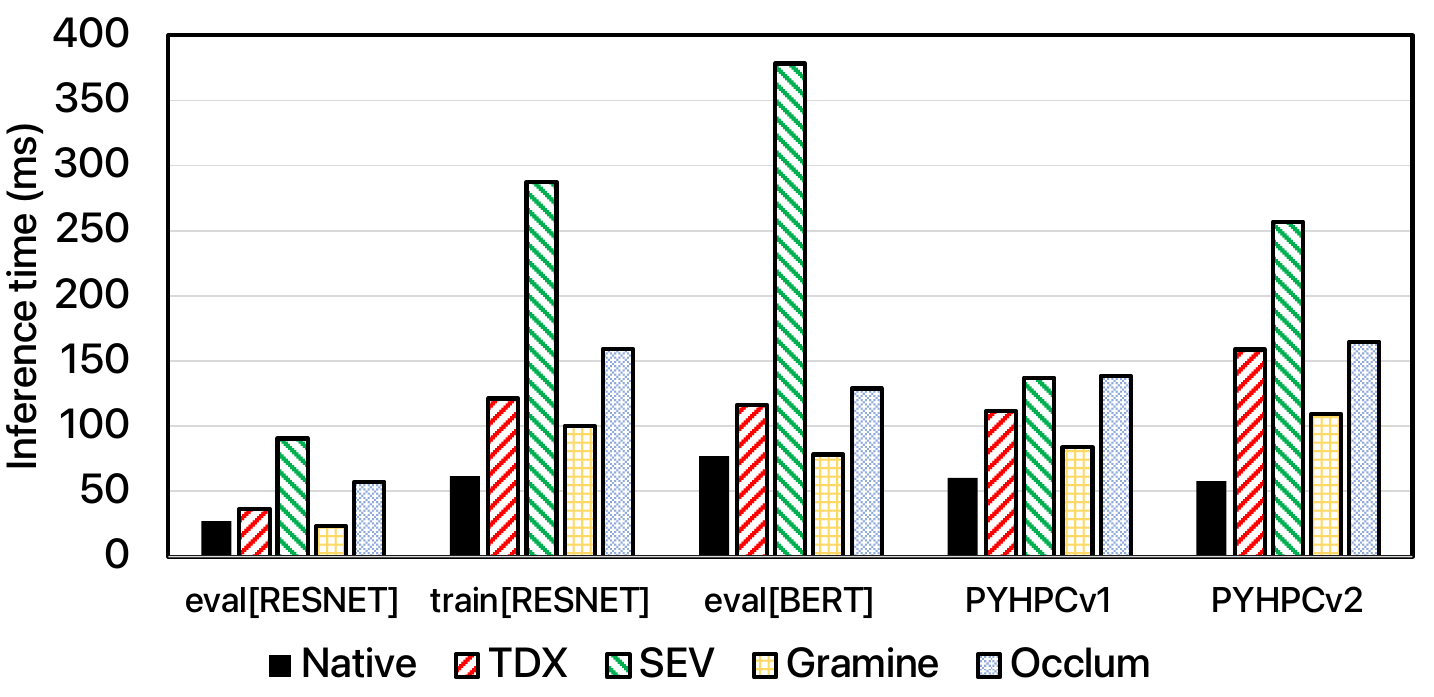}
        \label{fig:second}
    \end{subfigure}
    \hfill
    \caption{PyTorch Performance -- Inference Time}
    \label{fig:pytorch-perf}
\end{figure}

\begin{figure}
    \centering
    \includegraphics[scale=0.35]{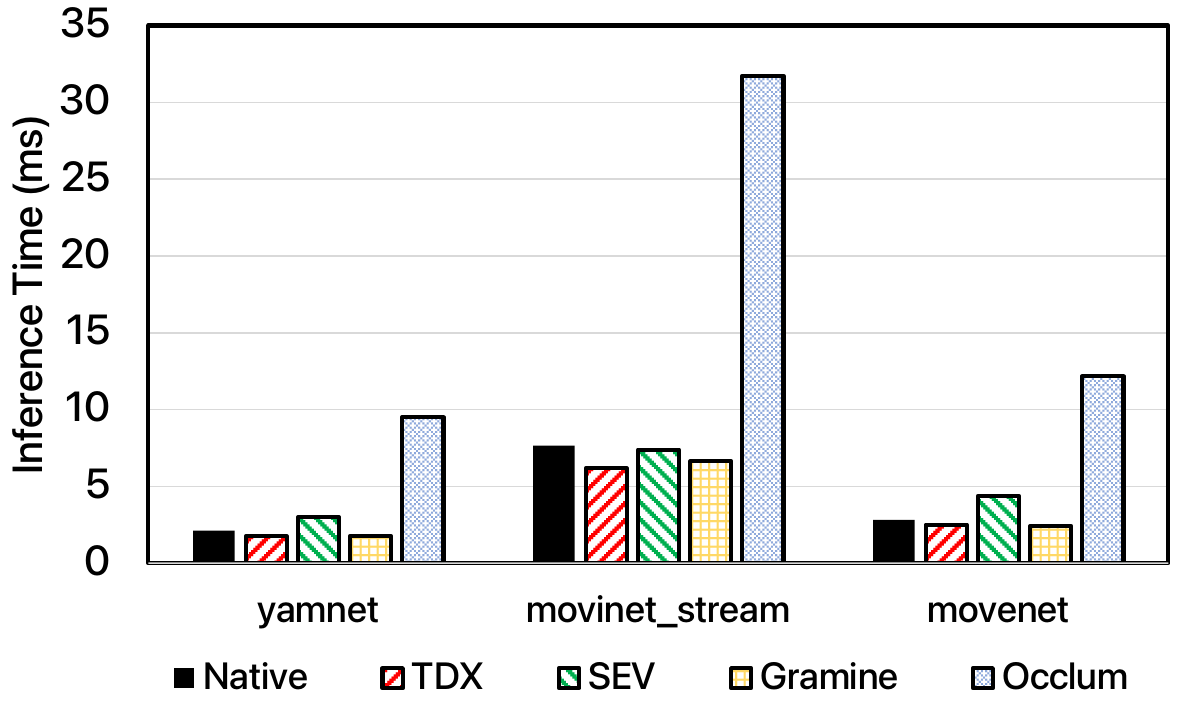}
    \caption{TensorFlow Performance -- Inference Time}
    \label{fig:tensorflow-perf}
\end{figure}

\subsection{CPU-Intensive Workloads}
Figures \ref{fig:pytorch-perf} and \ref{fig:tensorflow-perf} show bar graphs depicting the inference time of \emph{PyTorch} and \emph{TensorFlow} workloads, respectively. On the $x$-axis, we report the different benchmarks used for the evaluation. 
\\For what concerns \emph{PyTorch}, as expected the \textit{Native} environment consistently exhibits the shortest inference times across all benchmarks. \textit{Gramine-SGX} performs exceptionally well, often approaching the performance of the \textit{Native} environment. Unlike the other evaluations, \textit{Gramine-SGX} behaves better than \emph{TDX} across all benchmarks. \textit{Occlum} also exhibits better performance than usual, which is on the same level as TDX. SEV demonstrates significantly longer execution times, with approximately $40\%$ of this increase attributed to inherent CPU characteristic differences. 
\\In the case of \emph{TensorFlow}, the situation is different. TDX experienced a low inference time, which sometimes is even better than the \emph{Native} one. While \emph{Gramine-SGX} and \emph{SEV} reported a slightly higher execution time. \emph{Occlum} was the worst one with a higher inference time, up to $6\times$. 
\\What becomes evident from these findings is that Process-based TEEs tend to experience only minor performance degradation when subjected to CPU-intensive workloads. In certain instances, their performance closely matches that of VM-based TEEs, while in others, such as with \textit{Gramine-SGX}, they even surpass VM-based TEEs in terms of performance.

\begin{figure}
    \centering
    \includegraphics[scale=0.35]{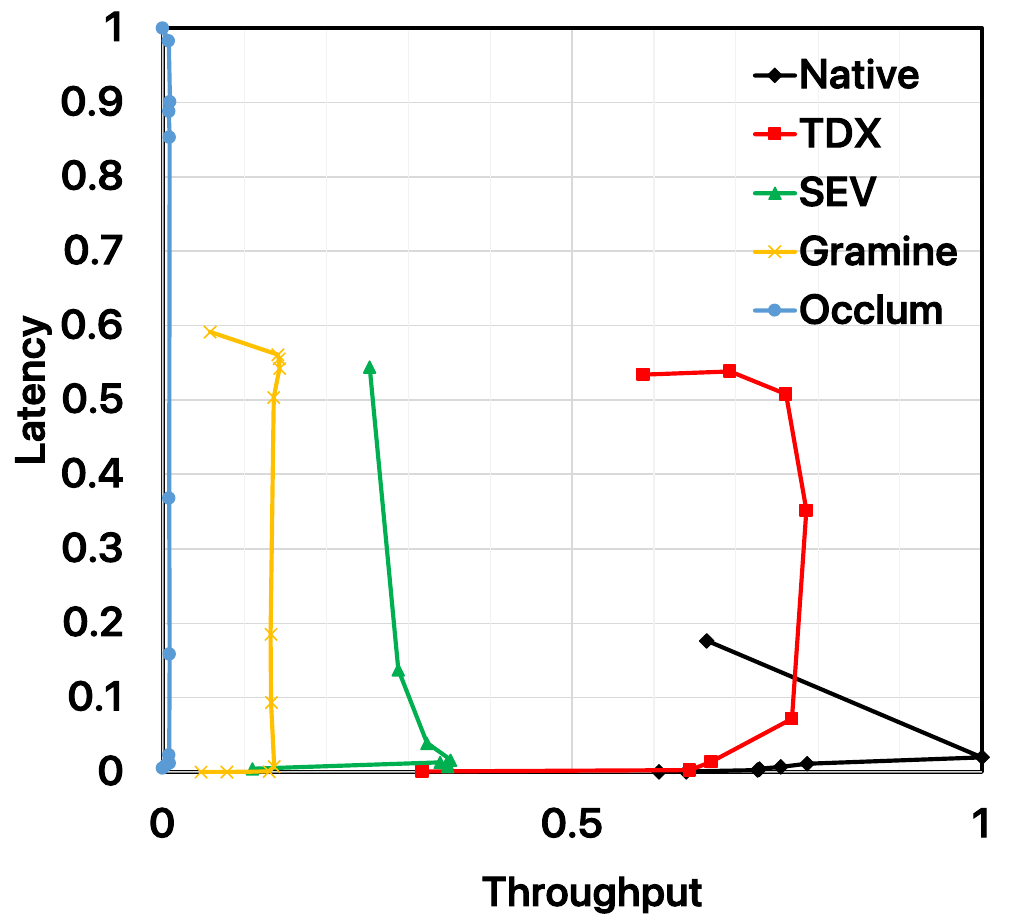}
    \caption{NGINX Performance -- Throughput vs Latency}
    \label{fig:nginx-perf}
\end{figure}

\begin{figure}
    \centering
    \includegraphics[scale=0.35]{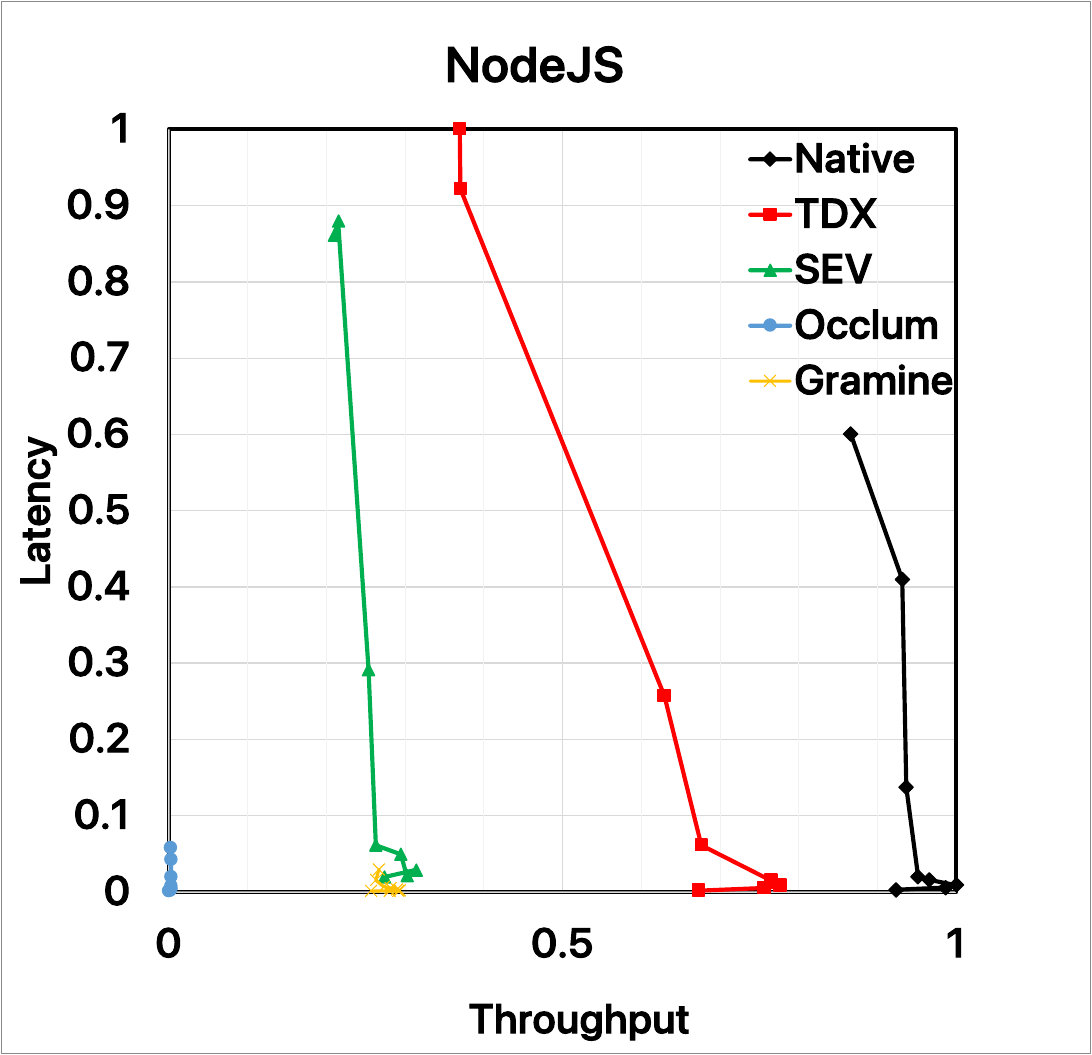}
    \caption{NodeJS Performance -- Throughput vs Latency}
    \label{fig:nodejs-perf}
\end{figure}

\begin{figure*}
    \centering
    \includegraphics[scale=0.4]{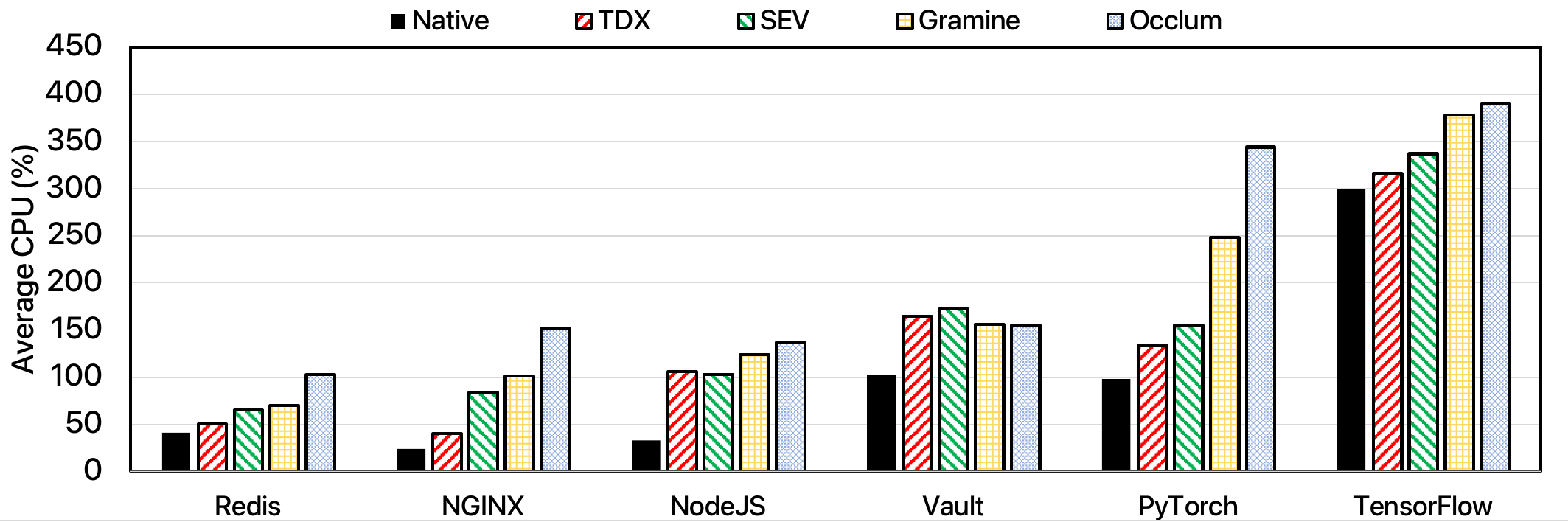}
    \caption{Average CPU Utilization}
    \label{fig:cpu-perf}
\end{figure*}

\subsection{I/O-Intensive Workloads}
Figure \ref{fig:nginx-perf} shows the performance of NGINX. The Native has the highest throughput. The latency remains the lowest across all throughputs. \textit{TDX} shows a significant drop in throughput compared to \textit{Native}, stabilizing just above 0.4. The latency is low when the throughput is below this point but rises sharply as the throughput increases. The average overhead of \textit{TDX} with respect to the \textit{Native} is $28.6\%$. SEV has a lower maximum throughput than \textit{TDX}. The latency is in the same range of \textit{TDX} at low throughputs but does not increase as sharply. Again, if we remove the impact of the AMD CPU, we can consider SEV $8.7\%$ slower than \textit{TDX}. \textit{Gramine-SGX} exhibits a similar pattern to \textit{SEV}, but with slightly higher latency across all throughput levels. Even in this case, \textit{Gramine-SGX} behaved better than \textit{Occlum-SGX} but still slower than the VM-based TEEs, reporting an overhead of $67.6\%$. 
\\Figure \ref{fig:nodejs-perf} compares the performance of the NodeJS workload. The \textit{Native} has the best performance as before, showing high throughput and low latency. While \textit{TDX} has a slightly higher latency than \textit{SEV}.
It is important to highlight that, for this particular workload, both \textit{Gramine-SGX} and \textit{Occlum-SGX} were not able to sustain the same amount of concurrent benchmark requests used for \textit{TDX}, \textit{SEV}, and \textit{Native} (in the range of $[10,8000]$). The webserver stuck after $500$ requests. Hence, a smaller range of requests has been used (i.e., $[10,500]$). This explains why the latency of \textit{Gramine-SGX} and \textit{Occlum-SGX} is extremely low compared to the other approaches.

\subsection{Resource Usage}
Besides evaluating the speed of our workloads, we are also interested in examining the CPU utilization across the various TEE approaches to better understand the trade-offs between security and computational efficiency.
\\In Figure \ref{fig:cpu-perf} we show the average CPU usage for the different workloads. As expected, the \textit{Native} runs exhibit the lowest CPU usage for all applications. \textit{TDX} shows competitive CPU usage, close to Native levels in most applications. AMD's \textit{SEV} consistently shows higher CPU usage compared to \textit{TDX} and sometimes even higher than \textit{Gramine} and \textit{Occlum}. Both process-based TEEs show variable CPU usage across different applications. \textit{Gramine} generally maintains moderate CPU usage, indicating a balanced performance. \textit{Occlum}, however, has notably high CPU usage with PyTorch, which could be indicative of less efficient handling of CPU-intensive workloads or a lack of optimization for this particular application. For Redis and NodeJS, all TEEs have relatively similar CPU usage, with TDX marginally outperforming the others. This suggests that for certain types of workloads, the choice of TEE might not significantly impact CPU efficiency. The PyTorch workload stands out with its high CPU usage in the Native environment, while TDX optimizes this usage considerably. Occlum, however, seems to struggle with this workload. TDX appears to be the most efficient in terms of CPU usage across a range of applications, closely followed by Gramine. SEV tends to have higher CPU overhead, while Occlum's performance is highly variable, performing well in some cases but not in others.

\section{Impact on Cloud Service Costs}
\label{servicecost}
\begin{figure}
    \begin{subfigure}{0.5\linewidth}
        \includegraphics[scale=0.34]{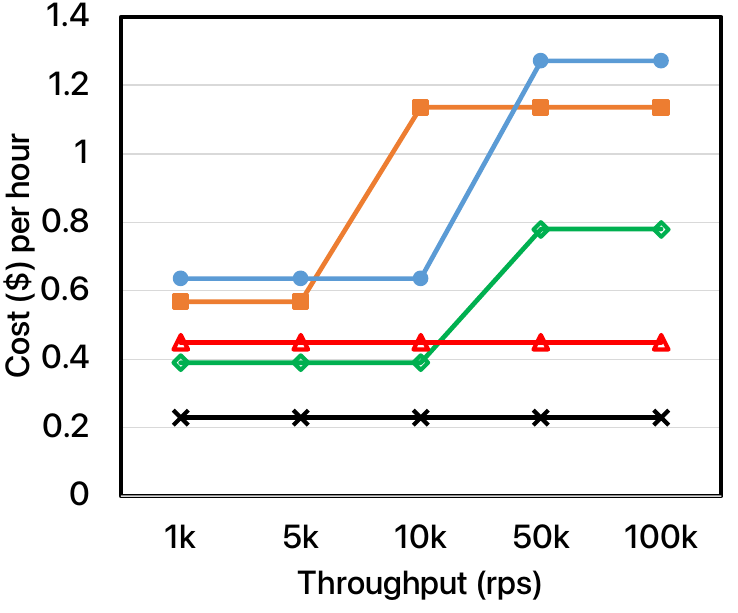}
        \caption{Redis}
        \label{subfig:redis}
    \end{subfigure}
    \begin{subfigure}{0.5\linewidth}
        \includegraphics[scale=0.34]{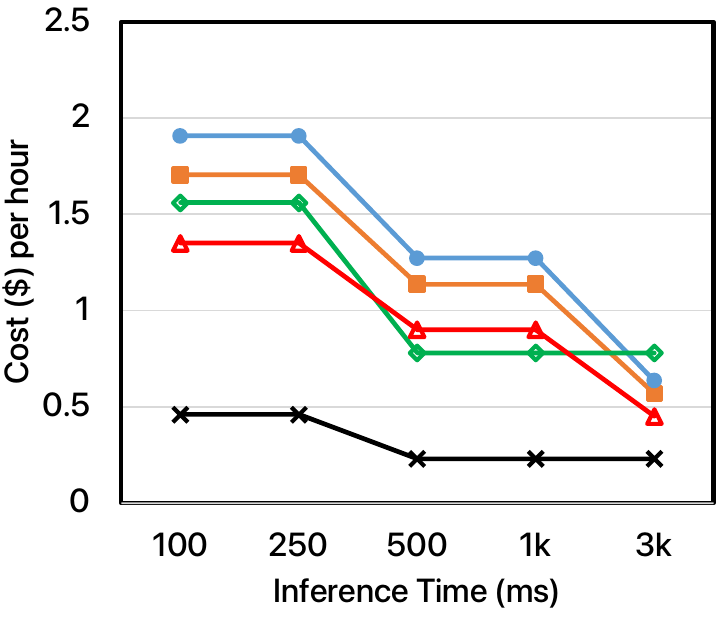}
        \caption{PyTorch}
        \label{subfig:pytorch}
    \end{subfigure}
    \begin{subfigure}{0.46\textwidth}
        \centering
        \includegraphics[scale=0.34]{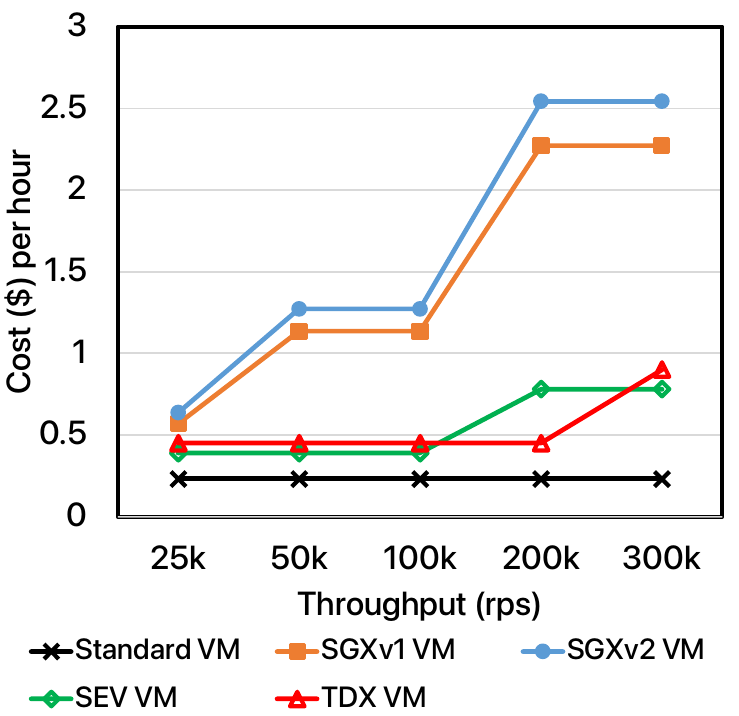}
        \caption{NGINX}
        \label{subfig:nginx}
    \end{subfigure}
    \caption{Costs of Cloud deployments}
    \label{fig:cloud-cost}
\end{figure}

This section aims to assess how different TEE solutions influence the costs associated with Cloud instances, particularly when striving to achieve predefined performance targets. To facilitate a precise estimation, we first collated the hourly rates from Azure Cloud for various machine types that meet the TEE hardware requirements. We started from the baseline configuration used during our experimental campaign, which consisted of virtual machines equipped with fixed disk size, 4 vCPUs, and 16 GB of vRAM. To evaluate how the variation of vCPUs and vRAM affects performance, we conducted experiments that adjusted these parameters. Notably, certain workloads, such as \textit{Redis}, showed no performance gains from increased core counts due to their single-threaded nature. In such instances, we deploy multiple instances in a clustered configuration under a load balancer to achieve the desired throughput. We point out that the increase of CPU cores in the Cloud VM configuration can only be done in a power of two. Regarding TDX, in the absence of official pricing at the time of this study, we provided an estimated cost. This estimate is based on the announced future availability of confidential TDX VMs on Azure, extrapolating from the current cost structure of similar VM services and accounting for the expected premium that TDX's enhanced security features would necessitate. Regarding SGX, we report costs for two typologies of instances having support for SGX \textit{v1} and \textit{v2} architectures. We selected representative workloads for each category --- i.e., \textit{Redis}, \textit{PyTorch}, \textit{NGINX} --- and varied benchmark performance targets in a range, which is based on the results obtained during our previous evaluations. We highlight that regardless of the CPU cores, the cost of SGX machines is the highest, followed by the TDX one, and lastly SEV.
\\Figure \ref{fig:cloud-cost} graphically shows the cost per hour of TEE-enabled VMs for each particular selected workload. Regarding \textit{Redis} (Figure \ref{subfig:redis}), it can be noticed that no increase of CPU cores --- thus increase of cost --- is needed for Standard and TDX VMs. Contrariwise, SGXv1 machines require more cores starting from $10k$ RPS. While, SGXv2 and SEV increase costs starting from $50k$ RPS. For what concerns \emph{PyTorch} (Figure \ref{subfig:pytorch}), we used as a benchmark the training of the BERT algorithm. It can be noticed that SEV is the one that costs more when the inference time must be lower than $3s$, and also it required much more cores when we targeted lower inference times. SGX VMs had a trend similar to the TDX one. This was expected given the performance results we obtained for CPU-intensive workloads.  Finally, regarding NGINX, the TDX machine doubles the cores only when a target of $300k$ RPS has to be achieved. While, for the SEV one, it happens with a $200k$ RPS target. SGX machines, instead, require a very high number of cores leading to significantly larger costs. Using $100k$ RPS, the situation gets better.

\section{Concluding Remarks}
\label{conclusion}
This paper features an in-depth comparative analysis -- covering cost, effort, security, and performance -- of some of the major solutions for transparent TEE protection of existing applications. We examined the performance and the cost differences of \textit{VM-based TEEs} (specifically, Intel TDX and AMD SEV) against \textit{Process-based TEEs} (i.e., Intel SGX) when used with runtimes such as \textit{Gramine-SGX} and \textit{Occlum-SGX}. The study provides decision-makers with insights useful for understanding which specific TEE solution best suits the requirements/constraints of a given setup. Our research demonstrates that for I/O- and Memory- intensive workloads the \textit{VM-based} TEEs are much better performing than \textit{Process-based} ones, while for CPU-intensive workloads, process-based TEEs emerge as a good option since the gain in terms of security comes at a lower cost of performance. Our findings indicate that TDX behaves better than SEV, a discrepancy that cannot be solely attributed to the intrinsic differences in Intel and AMD CPUs performance. Even after adjusting for potential CPU-related performance disparities, TDX does better in resource usage efficiency. In the area of process-based TEEs, \textit{Gramine-SGX} consistently outperforms \textit{Occlum-SGX} across all evaluated parameters, including resource consumption. However, our study also notes that process-based TEEs, while being a good compromise for CPU-bound tasks, tend to exhibit significantly higher resource usage overall. 
\\In terms of costs, in general, SGX deployments are the most expensive, followed by TDX, and then SEV. Using SGX or SEV with memory-intensive workloads requires more CPU cores, which results in higher costs (whereas this is not the case with TDX). For CPU-intensive workloads, the number of cores was increased with respect to a Standard VM for all types of VMs. For I/O-intensive ones, SGX VMs required a double level of cores due to the significant overhead suffered by the workloads.

\section*{Acknowledgments}
This project has received funding from the European Union's Horizon Europe Research and Innovation Programme under Grant Agreement No. 101070670 (ENCRYPT - A Scalable and Practical Privacy-preserving Framework). 
\\The work made in this paper was also funded by the European Union under NextGenerationEU. PRIN 2022 Prot. n.  202297YF75.
\\The authors would like to thank Alessandro De Crecchio for his valuable contribution to the experimental campaign.

\ifCLASSOPTIONcaptionsoff
  \newpage
\fi

\bibliographystyle{IEEEtran}
\bibliography{references}

\begin{thebibliography}{10}
\providecommand{\url}[1]{#1}
\csname url@samestyle\endcsname
\providecommand{\newblock}{\relax}
\providecommand{\bibinfo}[2]{#2}
\providecommand{\BIBentrySTDinterwordspacing}{\spaceskip=0pt\relax}
\providecommand{\BIBentryALTinterwordstretchfactor}{4}
\providecommand{\BIBentryALTinterwordspacing}{\spaceskip=\fontdimen2\font plus
\BIBentryALTinterwordstretchfactor\fontdimen3\font minus
  \fontdimen4\font\relax}
\providecommand{\BIBforeignlanguage}[2]{{%
\expandafter\ifx\csname l@#1\endcsname\relax
\typeout{** WARNING: IEEEtran.bst: No hyphenation pattern has been}%
\typeout{** loaded for the language `#1'. Using the pattern for}%
\typeout{** the default language instead.}%
\else
\language=\csname l@#1\endcsname
\fi
#2}}
\providecommand{\BIBdecl}{\relax}
\BIBdecl

\bibitem{survey}
\BIBentryALTinterwordspacing
L.~Coppolino, S.~D’Antonio, G.~Mazzeo, and L.~Romano, ``A comprehensive
  survey of hardware-assisted security: From the edge to the cloud,''
  \emph{Internet of Things}, vol.~6, p. 100055, 2019. [Online]. Available:
  \url{https://www.sciencedirect.com/science/article/pii/S2542660519300101}
\BIBentrySTDinterwordspacing

\bibitem{sgx}
\BIBentryALTinterwordspacing
F.~McKeen, I.~Alexandrovich, I.~Anati, D.~Caspi, S.~Johnson, R.~Leslie-Hurd,
  and C.~Rozas, ``Intel® software guard extensions (intel® sgx) support for
  dynamic memory management inside an enclave,'' in \emph{Proceedings of the
  Hardware and Architectural Support for Security and Privacy 2016}, ser. HASP
  '16.\hskip 1em plus 0.5em minus 0.4em\relax New York, NY, USA: Association
  for Computing Machinery, 2016. [Online]. Available:
  \url{https://doi.org/10.1145/2948618.2954331}
\BIBentrySTDinterwordspacing

\bibitem{gramine}
C.-C. Tsai, D.~E. Porter, and M.~Vij, ``Graphene-sgx: A practical library os
  for unmodified applications on sgx,'' in \emph{Proceedings of the 2017 USENIX
  Conference on Usenix Annual Technical Conference}, ser. USENIX ATC '17.\hskip
  1em plus 0.5em minus 0.4em\relax USA: USENIX Association, 2017, p. 645–658.

\bibitem{occlum}
\BIBentryALTinterwordspacing
Y.~Shen, H.~Tian, Y.~Chen, K.~Chen, R.~Wang, Y.~Xu, Y.~Xia, and S.~Yan,
  ``Occlum: Secure and efficient multitasking inside a single enclave of intel
  sgx,'' in \emph{Proceedings of the Twenty-Fifth International Conference on
  Architectural Support for Programming Languages and Operating Systems}, ser.
  ASPLOS '20.\hskip 1em plus 0.5em minus 0.4em\relax New York, NY, USA:
  Association for Computing Machinery, 2020, p. 955–970. [Online]. Available:
  \url{https://doi.org/10.1145/3373376.3378469}
\BIBentrySTDinterwordspacing

\bibitem{scone}
S.~Arnautov, B.~Trach, F.~Gregor, T.~Knauth, A.~Martin, C.~Priebe, J.~Lind,
  D.~Muthukumaran, D.~O'Keeffe, M.~L. Stillwell, D.~Goltzsche, D.~Eyers,
  R.~Kapitza, P.~Pietzuch, and C.~Fetzer, ``Scone: Secure linux containers with
  intel sgx,'' in \emph{Proceedings of the 12th USENIX Conference on Operating
  Systems Design and Implementation}, ser. OSDI'16.\hskip 1em plus 0.5em minus
  0.4em\relax USA: USENIX Association, 2016, p. 689–703.

\bibitem{sev}
D.~Kaplan, J.~Powell, and T.~Woller, ``{AMD Memory Encryption},'' AMD Developer
  Central, Advanced Micro Devices, Inc., pp. 1--12, Apr 2016, [Online].
  Available:
  \url{https://developer.amd.com/wordpress/media/2013/12/AMD_Memory_Encryption_Whitepaper_v7-Public.pdf}.

\bibitem{tdx}
``{Intel TDX},'' Intel Developer Reference, Nov 2023, [Online]. Available:
  \url{https://www.intel.com/content/www/us/en/developer/tools/trust-domain-extensions/overview.html}.

\bibitem{sahita}
R.~Sahita, D.~Caspi, B.~Huntley, V.~Scarlata, B.~Chaikin, S.~Chhabra,
  A.~Aharon, and I.~Ouziel, ``Security analysis of confidential-compute
  instruction set architecture for virtualized workloads,'' in \emph{2021
  International Symposium on Secure and Private Execution Environment Design
  (SEED)}, 2021, pp. 121--131.

\bibitem{felicitas}
\BIBentryALTinterwordspacing
F.~Hetzelt and R.~Buhren, ``Security analysis of encrypted virtual machines,''
  in \emph{Proceedings of the 13th ACM SIGPLAN/SIGOPS International Conference
  on Virtual Execution Environments}, ser. VEE '17.\hskip 1em plus 0.5em minus
  0.4em\relax New York, NY, USA: Association for Computing Machinery, 2017, p.
  129–142. [Online]. Available: \url{https://doi.org/10.1145/3050748.3050763}
\BIBentrySTDinterwordspacing

\bibitem{fei}
\BIBentryALTinterwordspacing
S.~Fei, Z.~Yan, W.~Ding, and H.~Xie, ``Security vulnerabilities of sgx and
  countermeasures: A survey,'' \emph{ACM Comput. Surv.}, vol.~54, no.~6, jul
  2021. [Online]. Available: \url{https://doi.org/10.1145/3456631}
\BIBentrySTDinterwordspacing

\bibitem{gottel-sevsgx}
C.~Göttel, R.~Pires, I.~Rocha, S.~Vaucher, P.~Felber, M.~Pasin, and
  V.~Schiavoni, ``Security, performance and energy trade-offs of
  hardware-assisted memory protection mechanisms,'' in \emph{2018 IEEE 37th
  Symposium on Reliable Distributed Systems (SRDS)}, 2018, pp. 133--142.

\bibitem{akram-sevsgx}
A.~Akram, A.~Giannakou, V.~Akella, J.~Lowe-Power, and S.~Peisert, ``Performance
  analysis of scientific computing workloads on general purpose tees,'' in
  \emph{2021 IEEE International Parallel and Distributed Processing Symposium
  (IPDPS)}, 2021, pp. 1066--1076.

\bibitem{mofrad-sevsgx}
\BIBentryALTinterwordspacing
S.~Mofrad, F.~Zhang, S.~Lu, and W.~Shi, ``A comparison study of intel sgx and
  amd memory encryption technology,'' ser. HASP '18.\hskip 1em plus 0.5em minus
  0.4em\relax New York, NY, USA: Association for Computing Machinery, 2018.
  [Online]. Available: \url{https://doi.org/10.1145/3214292.3214301}
\BIBentrySTDinterwordspacing

\bibitem{ccconsortium}
``{Common Terminology for Confidential Computing},'' Confidential Computing
  Consortium, Jan. 2024, [Online]. Available:
  \url{https://confidentialcomputing.io/wp-content/uploads/sites/10/2023/03/Common-Terminology-for-Confidential-Computing.pdf}.

\bibitem{8375bench}
``{Microbenchmark of Intel CPU 8375C},'' CPU Benchmarks, Dec. 2023, [Online].
  Available:
  \url{https://www.cpubenchmark.net/cpu.php?cpu=Intel+Xeon+Platinum+8375C+%40+2.90GHz}.

\bibitem{8480bench}
``{Microbenchmark of Intel CPU 8480},'' CPU Benchmarks, Dec. 2023, [Online].
  Available:
  \url{https://www.cpubenchmark.net/cpu.php?cpu=Intel+Xeon+Platinum+8375C+%40+2.90GHz}.

\bibitem{amdbench}
``{Microbenchmark of AMD EPYC 7763},'' CPU Benchmarks, Dec. 2023, [Online].
  Available:
  \url{https://www.cpubenchmark.net/cpu.php?cpu=Intel+Xeon+Platinum+8375C+%40+2.90GHz}.

\end{thebibliography}

\vspace{-10mm}
\begin{IEEEbiography}[{\includegraphics[width=1in,height=1.25in,clip,keepaspectratio]{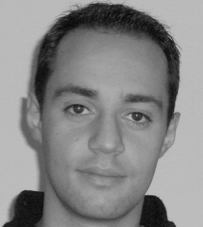}}]{Luigi Coppolino} PhD, is an Associate Professor at the University of Naples Parthenope, Italy.  His research activity mainly focuses on dependability of computing systems, critical infrastructure protection, and information security. He was the technical coordinator of the EC funded research project COMPACT and involved with key roles in several others.
\end{IEEEbiography}

\vspace{-10mm}
\begin{IEEEbiography}[{\includegraphics[width=1in,height=1.25in,clip,keepaspectratio]{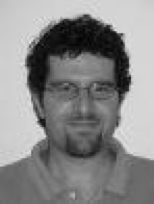}}]{Salvatore D'Antonio} is an Associate Professor at the University of Naples Parthenope, Italy. He is an expert in network monitoring, network security and critical infrastructure protection. He was the Coordinator of two EU research projects on critical infrastructure protection, namely INSPIRE and INSPIRE-INCO.
\end{IEEEbiography}

\vspace{-10mm}
% \begin{IEEEbiography}[{\includegraphics[width=0.6in,height=1.25in,clip,keepaspectratio]{Figures/vformicola}}]{Valerio Formicola} \scriptsize PhD, is  Post-Doc Researcher at the University Parthenope. His research is focused on data-driven methods for the dependability and security of distributed systems. He has been Co-PI for DOE projects in USA, and Research Fellow at the University of Illinois at Urbana-Champaign. 
% \end{IEEEbiography}

\begin{IEEEbiography}[{\includegraphics[width=1in,height=1.25in,clip,keepaspectratio]{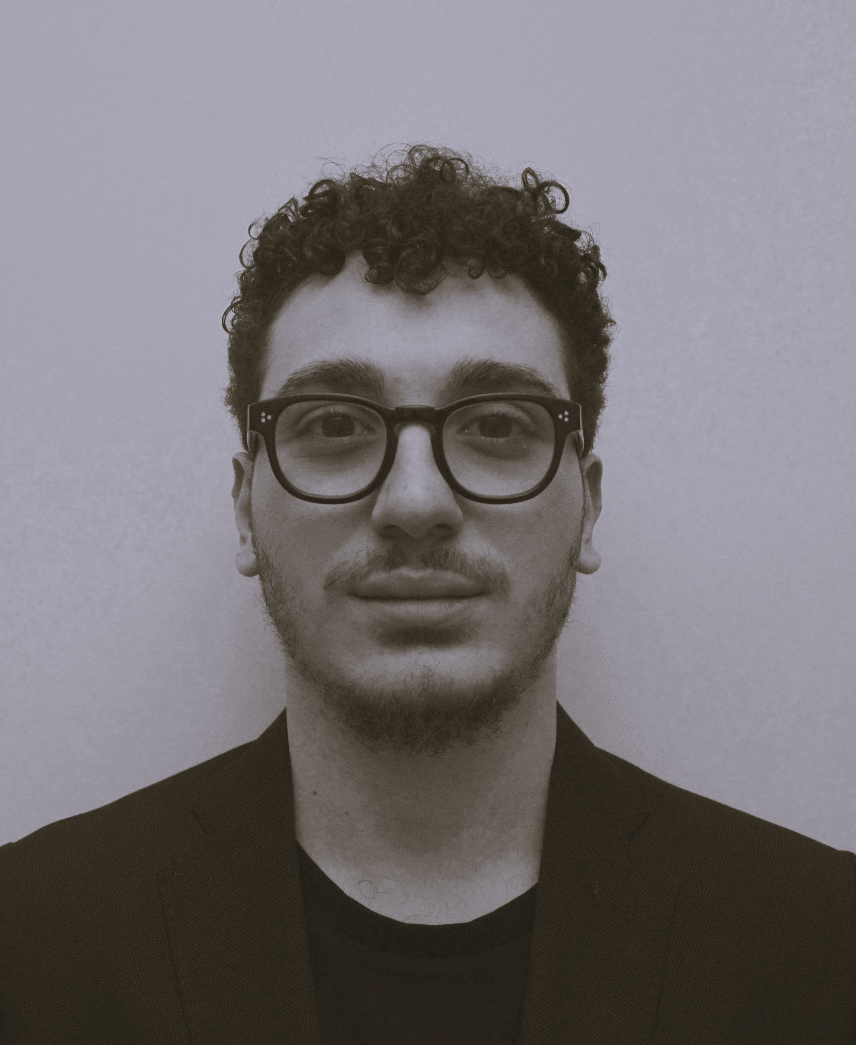}}]{Davide Iasio} is a Software Engineer at Trust Up srl. 
His background includes the development of microservices-based solutions for data protection in cloud environments and the management of cloud infrastructures.
\end{IEEEbiography}

\vspace{-10mm}

\begin{IEEEbiography}[{\includegraphics[width=1in,height=1.25in,clip,keepaspectratio]{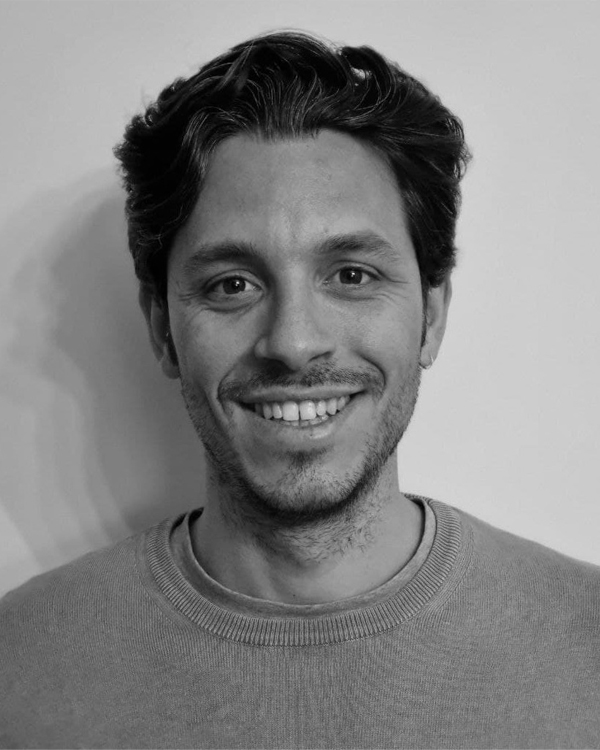}}]{Giovanni Mazzeo} PhD,
is an Assistant Professor at the Department of Engineering of the University of Naples Parthenope, Italy. His research field is the security and dependability of computer systems, with a particular focus on trusted computing. He was principal investigator of European research projects on IT security.
\end{IEEEbiography}
%\newpage

\vspace{-10mm}

\begin{IEEEbiography}[{\includegraphics[width=1in,height=1.25in,clip,keepaspectratio]{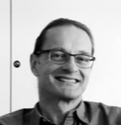}}]{Luigi Romano} PhD, is a Full Professor at the University of Naples Parthenope. His research interests are system security and dependability, with focus on Critical Infrastructure Protection. He has worked extensively as a consultant for industry leaders in the field of security- and safety-critical computer systems. He was one of the members of the ENISA expert group on Priorities of Research On Current and Emerging Network Technologies (PROCENT). 
\end{IEEEbiography}


%\vfill


\end{document}